\documentclass[aps,pra,showpacs,twoside,twocolumn,10pt]{revtex4-1}
\usepackage[colorlinks=true, citecolor=red, urlcolor=blue ]{hyperref}
\usepackage{epsfig,newlfont,amssymb,amsfonts,amsmath,bm,subfigure,palatino,mathtools,amsthm,braket,times,soul,enumitem,color}
\usepackage[normalem]{ulem}
\newcommand{\stkout}[1]{\ifmmode\text{\sout{\ensuremath{#1}}}\else\sout{#1}\fi}
\definecolor{magenta}{rgb}{1.0, 0.0, 0.56}

\begin{document}
 
 
 
 

\title{Heat current and entropy production rate in local non-Markovian quantum dynamics\\of global Markovian evolution}

\author{Ahana Ghoshal and Ujjwal Sen}
\affiliation{Harish-Chandra Research Institute, A CI of Homi Bhabha National Institute, Chhatnag Road, Jhunsi, Prayagraj 211 019, India}

\begin{abstract}
We examine the elements of the balance equation of entropy in open quantum evolutions and their response as we go from a Markovian to a non-Markovian situation. In particular, we look at the heat current and entropy production rate in the non-Markovian reduced evolution, as well as a Markovian limit of the same, experienced by one of two interacting systems immersed in a Markovian bath. The analysis naturally leads us to define a heat current deficit and an entropy production rate deficit, which are differences between the global and local versions of the corresponding quantities. The investigation leads, in certain cases, to a complementarity of the time-integrated heat current deficit and the relative entropy of entanglement between the two systems.
\end{abstract}

\maketitle

\section{Introduction}
\label{sec:intro}
In the past few decades, quantum thermodynamics has become an important area of research; in particular, its interface with  quantum information has been delved into. Thermodynamic laws
in the quantum regime have been introduced and scrutinized (see, e.g., \cite{Allahverdyan,Gemmer,Kosloff,Brand,Gardas,Gelbwaser,Misra,Millen,Vinjanampathy,Goold,Benenti,Binder,Deffner}). Quantum thermal devices have been designed (see, e.g., \cite{Palao,Feldmann,Levy1,Levy,Uzdin,Clivaz,Mitchison}), and  advantages over 
their classical counterparts have been investigated (see, e.g., \cite{Geva,Feldmann1,Ng,Niedenzu,Xu2}). These developments have contributed towards blurring the boundary of quantum thermodynamics with the theory of open quantum systems.
Two broad and important branches in this arena are those of  equilibrium  and nonequilibrium thermodynamics. Nonequilibrium situations are of course easily available in nature 
\cite{Lebon,Groot,Demirel,Kleidon,Ottinger,Holian,Evans,Hoover,Chernov,Gallavotti}. 
\par
The entropy production rate (EPR) provides evidence for nonequilibrium phenomena  \cite{Nicolis} (see also \cite{Thomsen,Yang,Ruelle,Daems,Bag,Lendi}). 
For an open quantum system
\cite{Petruccione,Alicki,Rivas,Lidar},
the EPR is a fundamental quantity which can provide 
important information about its thermal and steady-state properties.
The balance equation of entropy for a system immersed in a bath
\cite{Groot,Lebowitz,Spohn} reads
\begin{equation*}
    \frac{dS}{dt} + J
= \sigma,
\end{equation*}
where $S$ is the entropy of the system, 
$J$ quantifies the flow of entropy (heat) from the system to the bath, and 
$\sigma$ is the source term quantifying the entropy production rate of the system. These quantities will be defined more carefully later. 
The entropy production rate can be defined, for Markovian evolutions, as the negative derivative of the relative entropy distance from 
the canonical equilibrium state of the system, 
and it is also a measure of dissipativity of the 
dynamics
\cite{Lebowitz,Spohn}. 
The positivity of the EPR for Markovian quantum  processes is referred to as Spohn's theorem.
The definition, using the derivative of relative entropy, of EPR can be altered for 
non-Markovian evolutions 
and if used therein 
may 
lead
to a negative EPR \cite{Bylicka,Marcantoni,Popovic,Bhattacharya,Strasberg,Naze,Bonanca}.
The entropy production rate has been  measured experimentally in  \cite{Tietz,Jiang1,Rossi}.
See \cite{Alicki1,Turitsyn,Andrae,
Vollmer,Breuer,Esposito,Abe,Bandi,Deffner1,Gardas,Salis,Xing,Carlen,Xu1,Kawazura,Hase,Kuzovlev,Banerjee,Yu,Xu,Wang,Chen,Taye,Zhang,Ploskon,Jaramillo,Santos,Zeng,Busiello,Dixit,Kanda,Seara,Lee,Wolpert,Goes,Przymus,Budhiraja,Zicari,Kappler,Gibbins,Li,Horowitz}
for work 
on 
entropy production and entropy production rate
in nonequilibrium situations. Entropy production in a quantum impurity model has been studied in \cite{Mehta}. 
Microscopic expressions of entropy production and the entropy production rate for an open quantum system weakly coupled to a heat reservoir were derived in \cite{Lutz} (see also 
\cite{Cai}). 
\par

In this paper we look at the heat current and entropy production rate in a particular non-Markovian situation. Precisely, we consider a system consisting of a pair of quantum two-level systems (TLSs), generally interacting with each other, and immersed in a bath. The interaction of the entire system of two qubits with the bath is Markovian. However, when considered separately, the time evolution of any of the qubits is non-Markovian.  We subsequently consider the heat current deficit, defined as the difference between the global and local heat currents, with the global one being the heat current of the entire two-qubit system and the local one being the sum of those of the single qubits. In parallel, we also consider the entropy production rate deficit. These  deficits 
are analogous to
the ones for other well-known quantities. As examples we mention the deficit for entropy  that leads to the definitions of classical and quantum mutual information \cite{Cover, Zurek} and the deficit for work done by heat engines that leads to the definition of  quantum work deficit \cite{Oppenheim, Horodecki, Sen, Modi, Bera}. 
To underline the nontrivial effects produced due to the non-Markovianity in the evolution, we separately examine the case when the evolution of any one qubit is considered as oblivious of the existence of the other one, with the situation being referred to as the lone-qubit Markovian limit. In the non-Markovian case, we then analyze two separate classes of instances, viz., (i) 
when the two qubits are interacting via an Ising interaction with a parallel field
 or 
 (ii) when the two qubits are interacting 
through an \(XYZ\) spin exchange in a field. 
Both classes contain the case when the two qubits are not interacting as a limiting case, which is equivalent to the lone-qubit Markovian limit.
Class (ii) also contains the case when the two qubits are interacting by means of an \(XY\) spin exchange with a transverse field.
%
%
Evidence for non-Markovianity in these instances, for the single-qubit evolutions, is obtained by using the approaches of Breuer et al.~\cite{Piilo} and Rivas et al.~\cite{Huelga}. 
We go on to define the time-integrated heat current deficit and refer to it as the total heat current deficit. Further, it turns out that it has a complementary relation, in certain cases, with the entanglement~\cite{Baltic} in the two-qubit evolved state.

It is important to mention here a non-Markovian evolution in the reduced system dynamics of a two-qubit single-bath system, studied in Ref. \cite{Popovic} (see also Refs. \cite{Campbell,Cimmarusti,Raja} in this regard). In \cite{Popovic}, Popovic et al. studied correlations between the qubits and the entropy production rate and showed that the negativity of the EPR is a signature of non-Markovian evolution. They assumed that only one qubit experiences the effect of the environment and the other qubit is acting as an auxiliary which is strongly coupled with the first qubit and accordingly the authors set the master equations. In contrast, in our case, both the qubits are interacting with the environment and we are considering the dynamics of one qubit in the presence of the other; this leads to  the non-Markovian scenario in the reduced dynamics.

Let us comment here on the choice of the non-Markovianity detectors. There are a  number of conceptualizations of non-Markovianity and correspondingly several criteria have been proposed. However, it is well known that the different conceptualizations often do not agree with each other. Moreover, the corresponding criteria are often not necessary and sufficient, but only sufficient. Furthermore, several criteria are not tractable analytically or numerically. There is at present no known criterion that is necessary and sufficient as well as tractable. This is the reason we will check for the presence of non-Markovianity using two conceptually different criteria.
We will also find that the initial state used in an evolution affects the detection of non-Markovianity. This is probably expected. The situation is similar to entanglement generation using global operations (e.g., global unitaries), where the global operation does not lead to entanglement generation for arbitrary input states. 
Non-Markovianity detection criteria are also similar to entanglement detection criteria in that there is as yet no tractable criterion for entanglement detection that is both necessary and sufficient.


The remainder of the paper is presented as follows. In Sec. \ref{sec2} we discuss the Markovian master equation and the corresponding Lindblad operators for a pair of two-level systems interacting with a common bath. We mention here the two types of local dynamics of the single-qubit systems that we will be considering in this paper, viz., the lone-qubit Markovian limit and the reduced non-Markovian dynamics. 
In Sec. \ref{sec3} we discuss the EPR and heat current, their deficits, and the corresponding total deficits within the lone-qubit Markovian limit. 
In Secs. \ref{sec4} and \ref{sec5} 
we consider the reduced non-Markovian dynamics, of which we examine several cases, differentiated by the type of interaction between the two qubits of the system. In Sec. \ref{sec4} we study the EPR and heat current, their deficits, and the corresponding total deficits in the non-Markovian cases. 
Non-Markovianity of the evolutions is demonstrated by using the approaches of Breuer et al. and Rivas et al. in Sec. \ref{sec5}.
The complementarity with entanglement of the total heat current deficit in the lone-qubit Markovian limit and the non-Markovian case is taken up in Sec. \ref{sec6}. A summary is presented in Sec. \ref{sec7}.

\section{Two paths to local dynamics for  global Markovian evolution}
\label{sec2}
\begin{figure}
\includegraphics[height=4cm,width=7cm]{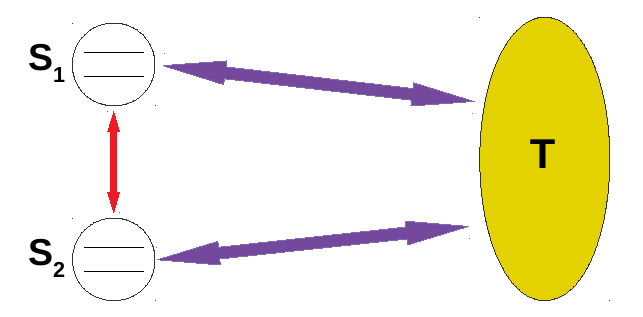}
\caption{Two qubits and a bath. We schematically depict here the situation considered in this paper. A TLS, $S_1$, is interacting with a bath at temperature \(T\). Another TLS, $S_2$, is also interacting with the same bath. The two-level systems are also interacting with each other.
}
\label{fig1}
\end{figure}

We consider a pair of two-level systems, with the  density matrix of the whole two-qubit system being  denoted by $\rho_{12}$, interacting with a common thermal bath $B$ at temperature $T$, as schematically depicted in Fig. \ref{fig1}. 
We consider the dynamics of the two-qubit system within the Markovian approximation, so that the interaction strength between the 
components of the bath is being assumed to be stronger than the system-bath interaction strength.
The heat current and entropy production rate of a single system 
are well studied \cite{Petruccione,Alicki,Spohn,Lebowitz} and therefore can be carried over directly to the case when we consider the entire two-qubit system.
We however wish to study the heat currents and entropy production rates of the individual qubits, so their individual dynamical equations are necessary. To obtain them, we consider two different scenarios. The first is in the extreme case when the dynamical equation of any one of the qubits is considered oblivious of the other. 
In this case, therefore, the local dynamics of any of the qubits is again Markovian. 
The second case is when this presumption of ignoring the other qubit when considering the dynamics of any of the qubits is not conceded. The first case is taken up in Sec. \ref{sec3} and the second in Secs. \ref{sec4} and \ref{sec5}. We refer to the first case as the lone-qubit Markovian limit and the second as simply the case of non-Markovian dynamics.

For completeness, we briefly present here the Markovian dynamical equation of the system of two qubits. Initially, the system-bath density matrix is a product state between the system and the bath, and is 
defined as 
$\rho(0)=\rho_{12}(0) \otimes \rho_B$, where $\rho_{12}(0)$ is a two-qubit pure state and $\rho_B$ is the initial state of the thermal bath. 
We have chosen the thermal bath as a collection of harmonic oscillators. The Hamiltonian of the system-bath configuration is 
\begin{equation}
H=H_{12}+H_{\text{bath}}+V_{12},
\label{Hamiltonian}
\end{equation}
where $H_{12}$ and $H_{\text{bath}}$ are the Hamiltonians of the system pair and the harmonic-oscillator bath, respectively, 
and $V_{12}$ denotes the interaction between the system pair and the bath. The three parts of $H$ are given by
\begin{eqnarray}
\label{H12}
&&H_{12}=\sum_{i}\frac{\hbar\omega_i}{2}\sigma_z^i+g H_{\text{int}}\equiv\sum_i H_i+g H_{\text{int}},\\
&&H_{\text{bath}}= \int_0^{\omega_{k_{\text{max}}}} \hbar \tilde{\omega} d\omega_{k} a_{\omega_{k}}^{\dagger} a_{\omega_{k}},\\
&&V_{12}=\sum_{i}\int_0^{\omega_{k_{\text{max}}}} \hbar \sqrt{\tilde{\omega}} d\omega_{k} h(\omega_{k})(\sigma_+^i a_{\omega_{k}}+\sigma_-^i a_{\omega_{k}}^{\dagger}).
\end{eqnarray}
Here $i=1,2$; $H_{\text{int}}$ is the interaction Hamiltonian between the two TLSs; $g$ is the corresponding coupling constant having the unit of energy;
$\tilde{\omega}$ is a constant having the unit of frequency; $a_{\omega_{k}}^{\dagger} (a_{\omega_{k}})$ is the bosonic creation (annihilation) operator of the harmonic-oscillator of the $k$th mode of the bath, being in units of $\frac{1}{\sqrt{\omega_{k}}}$ and obeying the commutation relation $[a_{\omega_{k}},a_{\omega_{k}^\prime}^\dagger]=\delta(\omega_{k}-\omega_{k}^\prime)$; and $h(\omega_{k})$ tunes the system-bath coupling strength and is a function of $\omega_{k}$. Precisely, $\tilde{\omega}h^2(\omega_{k})=J(\omega_{k})$, where $J(\omega_{k})$ is the spectral function of the harmonic-oscillator bath. We will consider here 
the spectral density function as Ohmic, so
$J(\omega_{k}) \propto \omega_{k}$. Thus $J(\omega_{k})=\lambda \omega_{k}$, where $\lambda$ is a unit-free constant. 
For the local TLS Hamiltonian, \(H_i\), of the $i$th qubit, the ground state having energy $-\frac{\hbar \omega_i}{2}$ is expressed by $\ket{1}$ and the excited state having energy $\frac{\hbar \omega_i}{2}$ is expressed by $\ket{0}$. The system-bath interaction Hamiltonian, $V$, can be decomposed as $V=\sum_k A_k \otimes B_k$, where $A_k$ and 
$B_k$ are system and bath operators respectively. Let $\ket{j}$ and $\ket{l}$ be two nondegenerate eigenvectors of the system Hamiltonian, $H_{12}$, corresponding to energy eigenvalues $\hbar\omega_{j}$ and $\hbar\omega_{l}$ respectively. Thus the transition energy associated with the transition from $\ket{j}$ to $\ket{l}$ is given by $\hbar\omega=\hbar\omega_l-\hbar\omega_j$. The number of $\omega$'s depends on the number of possible transitions (see \cite{Petruccione, Rivas} for more details).
The Born-Markov master equation for this TLS pair interacting with the thermal bath is 
given
by
\begin{equation}
\frac{d\rho_{12}(t)}{dt}=\mathcal{L}_{12}(\rho_{12})\equiv-\frac{i}{\hbar}[H_{12}+H_{{\text{LS}}},\rho_{12}(t)]+\mathcal{D}_{12}(\rho_{12}(t)),
\label{dyn_eq}
\end{equation}
where $\mathcal{D}_{12}(\rho_{12}(t))$ is the dissipative term which can be expressed as
\begin{eqnarray}
\mathcal{D}_{12}(\rho_{12}(t))=
\sum_{\omega}\gamma(\omega)\Big[A_{12}(\omega)\rho_{12}(t)A_{12}^{\dagger}(\omega)\nonumber\\-\frac{1}{2}\left\{A_{12}^{\dagger}(\omega)A_{12}(\omega),\rho_{12}(t)\right\}\Big],
\label{diss}
\end{eqnarray} 
where
\begin{equation}
\gamma(\omega)=
\begin{cases}
2\pi J(\omega)[\bar{n}(\omega)+1], & \omega > 0\\
2\pi J(|\omega|)\bar{n}(|\omega|), & \omega < 0.
\end{cases}
\end{equation} 
Here, $\bar{n}(\omega)=\frac{1}{e^{\hbar\omega/ k_{B}T}-1}$ is the Bose-Einstein distribution, with $k_B$ being the Boltzmann constant. In addition, $H_{\text{LS}}$ is the Lamb shift Hamiltonian,
whose effects, in the weak-coupling regime, i.e., for $\gamma \ll \{\omega_i,g\}$, usually satisfied in the optical quantum regime,
can be neglected 
\cite{Correa}. The 
Lindblad operators satisfy
\begin{equation}
\sum_{\{\omega\}} A_{12}(\omega)=\sum_{i=1}^2(\sigma_+^i+\sigma_-^i)
\end{equation}
and 
are given by
\begin{equation}
A_{12}(\omega)=\sum_{\omega=\hbar(\omega_{l}-\omega_{j})} \ket{j}\bra{j}\sum_{i=1}^2(\sigma_+^i+\sigma_-^i)\ket{l}\bra{l}.
\label{lindblad} 
\end{equation}
These are the eigenoperators of the system Hamiltonian obeying the commutation $[H_{12},A_{12}(\omega)]=-\omega A_{12}(\omega)$.\par
\begin{figure}
\includegraphics[height=4cm,width=7cm]{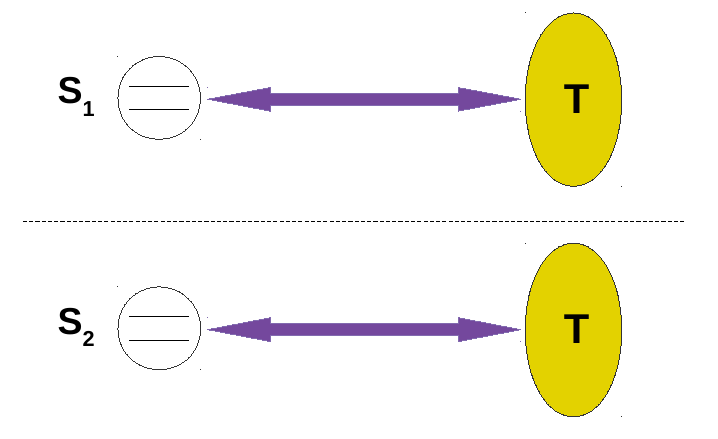}
\caption{Lone-qubit Markovian limit. Two TLSs are interacting with a bath at temperature $T$. It is assumed that the local dynamics of either one of them can be considered independently of the other. This scenario is considered in Sec. \ref{sec3} and in the first part of Sec. \ref{sec6}. 
Even though there is just a single bath in the set-up, it is placed twice in the schematic to underline the assumption that it is interacting with each of the TLSs while being oblivious to the presence of the other TLS.}
\label{fig2}
\end{figure}

\section{EPR, heat current and heat current deficit in lone-qubit Markovian limit}
\label{sec3}

In the present and succeeding sections, we consider the simple scenario where the two TLSs are interacting with a thermal bath at temperature \(T\) and the local dynamics of each of the TLSs can be considered independently of the other. This is schematically 
%
%
shown in Fig. \ref{fig2}.

The initial density matrix of the $i$th TLS is the state obtained by a trace on the other system, i.e., $\rho_{1}(0)=\text{Tr}_{2}[\rho_{12}(0)]$, and similarly for the other TLS. We are taking the partial trace only at the initial time; and after that the two systems are evolving independently.  
The Hamiltonian for the $i$th system is given by
\begin{equation}
H_{i}=\frac{\hbar \omega_i}{2}\sigma_z^i
\end{equation}
and the interaction between the system and bath is expressed as
\begin{equation}
V_i=\int_0^{\omega_{k_{\text{max}}}} \hbar d\omega_{k} h(\omega_{k})(\sigma_+^ia_{\omega_{k}}+\sigma_-^ia_{\omega_{k}}^{\dagger}).
\end{equation}
The dynamical equation for this case is given by Eqs. (\ref{dyn_eq}) and (\ref{diss}), only with $12$ being replaced by $i$. The system operators satisfy 
\begin{equation}
\sum_{\{\omega\}} A_{i}(\omega)=\sigma_+^i+\sigma_-^i,
\end{equation}
with $[H_{i},A_{i}(\omega)]=-\omega A_i(\omega)$.
 The Lindblad operators are given by
\begin{equation}
A_{i}(\omega)=\sum_{\omega=\hbar\omega_{\tilde{l}}-\hbar\omega_{\tilde{j}}} \ket{\tilde{j}}\bra{\tilde{j}}(\sigma_+^i+\sigma_-^i)\ket{\tilde{l}}\bra{\tilde{l}}, 
\label{lindblad-eTa-aar-use-hoini} 
\end{equation}
where $\ket{\tilde{j}}$ and $\ket{\tilde{l}}$ are the nondegenerate eigenvectors of $H_{i}$ corresponding to energy eigenvalues $\hbar\omega_{\tilde{j}}$ and $\hbar\omega_{\tilde{l}}$, respectively. There are two possible transition energies and two corresponding Lindblad operators. 
They are given by
\begin{equation}
\omega=\pm\hbar\omega_i, \quad A_i(\omega)=\sigma_{\mp}^i.
\end{equation}
Note that $A_i(-\omega)=A_i^{\dagger}(\omega)$. 

Let us digress in this paragraph to a generic system 
and consider an arbitrary dynamical map $V(t)$ for a quantum system governed by the system Hamiltonian \(H_{\text{sys}}\). 
The dynamical map is expressed as $V(t)=\exp(\mathcal{L}_{\text{sys}}t)$, where the operator $\mathcal{L}_{\text{sys}}$ appears in the dynamical equation,
\begin{equation}
\frac{d\rho_{\text{sys}}(t)}{dt}=\mathcal{L}_{\text{sys}}\rho_{\text{sys}}(t).
\label{dyn_equ}
\end{equation} 
The canonical equilibrium state of the system is given by $\rho_{\text{th}}=\exp(-\beta H_{\text{sys}})/\text{Tr}[\exp(-\beta H_{\text{sys}})]$, where 
$\beta=1/(k_BT)$.
In nonequilibrium thermodynamics, the relation between the entropy production rate and heat current of an open system is given by \cite{Groot,Lebowitz,Spohn,Petruccione}  
\begin{equation}
\frac{dS}{dt}+J=\sigma.
\label{relation}
\end{equation}
This is 
an expression of the second law of thermodynamics. It is also similar in spirit to the continuity equations in fluid mechanics \cite{Thomson}, quantum mechanics \cite{Mathews}, etc. Here $S$ is the von Neumann entropy of the open system and is defined by 
\begin{equation}
S(\rho_{\text{sys}})=-k_B\text{Tr}[\rho_{\text{sys}}\ln(\rho_{\text{sys}})]=-k_B\sum_{j}\lambda_j\ln(\lambda_j),
\label{entropy}
\end{equation} 
where the $\lambda_j$ are the eigenvalues of the density matrix $\rho_{\text{sys}}$ and 
$J$ is the entropy flux, defined as the amount of entropy exchanged per unit time between the open system and the environment \cite{Petruccione}. This entropy flux can also be referred to as the heat current.
For $J>0$, heat flows from the system to the environment, and $J<0$ means the opposite. Alternatively, it can be said that $J$ is produced for the changes of internal energy due to dissipative effects. Thus $J$ can be defined as
\begin{equation}
J=-\frac{1}{T}\text{Tr}[H_{\text{sys}}\mathcal{D}_{\text{sys}}(\rho_{\text{sys}})]=-\frac{1}{T}\text{Tr}[H_{\text{sys}}\mathcal{L}_{\text{sys}}(\rho_{\text{sys}})],
\label{hc}
\end{equation}
with \(\mathcal{D}_{\text{sys}}\) representing the dissipative term in the dynamics.
Using the definitions of $S$ and $J$, i.e., Eqs. (\ref{entropy}) and (\ref{hc}), respectively, one can show that
\begin{eqnarray}
\nonumber
&&\frac{dS}{dt}=-k_B\text{Tr}[\mathcal{L}_{\text{sys}}(\rho_{\text{sys}})\ln(\rho_{\text{sys}})]\\
\text{and} \quad &&J=k_B\text{Tr}[\mathcal{L}_{\text{sys}}(\rho_{\text{sys}})\ln(\rho_{\text{th}})].
\label{eq:S,J}
\end{eqnarray}
Here $\sigma$ is the entropy source strength and is referred to as the 
entropy production rate, i.e., the amount of entropy produced in a unit of time as a result of irreversible processes. Using Eq. (\ref{eq:S,J}), $\sigma$ can be expressed as the negative time derivative of the relative entropy distance of $\rho_{\text{sys}}(t)$ from the canonical equilibrium state $\rho_{\text{th}}$, as shown in \cite{Lebowitz,Spohn};
\begin{equation}
\sigma=-\frac{d}{dt}S(\rho_{\text{sys}}||\rho_{\text{th}}),
\label{sigma}
\end{equation}
where $\rho_{\text{sys}}(t)=V(t)\rho(0)$, with \(\rho(0)\) the initial state of the evolution, and where we have made the assumption of Markovian dynamics, i.e., (i) the dynamical map $V(\cdot)$ is completely positive, (ii) $V(\cdot)$ is trace preserving, (iii) divisibility is always satisfied, i.e., $V(t+s)=V(t)V(s)$ for any intermediate times $t$ and $s$, 
and (iv) $\lim_{t \rightarrow 0_+} || V(t)\rho-\rho ||_1=0$ for all $\rho$. 
The relative entropy, \(S(\varrho || \varsigma)\), between the density matrices \(\varrho\) and \(\varsigma\) is defined as \(k_B\mbox{Tr}(\varrho \ln \varrho - \varrho \ln \varsigma)\). 
The negative time derivative of the relative entropy is a convex function and is always positive (\(\sigma \geq 0\)) for a Markovian evolution.
The positivity of the entropy production rate for Markovian evolutions is referred to as Spohn's theorem \cite{Spohn}. 
\par 
The balance relation between entropy production rate and  heat current 
is given, for \(a=12\), by
\begin{equation}
\frac{dS(\rho_{a})}{dt}+J_{a} = \sigma_{a},
\end{equation} 
where
\begin{equation}
\label{eq:EPR_m}
\sigma_{a}=-\frac{d}{dt}S(\rho_{a}(t)||\rho_{{\text{th}}_{a}}), \quad  J_{a}=-\frac{1}{T}\text{Tr}[H_{a}\mathcal{D}(\rho_{a})], \end{equation}
with $\rho_{\text{th}_{a}}$ the canonical equilibrium state for the two-qubit system interacting with the common bath at temperature \(T\), with the 
system Hamiltonian being given by Eq. (\ref{Hamiltonian}).
For the lone-qubit Markovian limit, there is simply a doublet of the above relations for \(a=1\) and \(a=2\). 


We wish to quantify the amount of heat current of the entire two-qubit system that is not accounted for by the local heat currents. To this end, we define the 
heat current deficit  
as the following difference:
\begin{equation}
\Delta J=J_{12}-J_1-J_2.
\end{equation}
The time integral of the heat current deficit can be called the total heat current deficit and is given by
\begin{eqnarray}
\nonumber
\int_0^t \Delta J dt&=&\int_0^t (J_{12}-J_1-J_2) dt\\
\nonumber&=&\int_0^t dt \Big[(\sigma_{12}-\sigma_1-\sigma_2)\\&&-\left(\frac{dS(\rho_{12})}{dt}-\frac{dS(\rho_1)}{dt}-\frac{dS(\rho_2)}{dt}\right)\Big].
\end{eqnarray}
Up to an additive quantity related to the quantum mutual information \cite{Cover, Zurek},
\begin{equation}
  I(\rho_{12}) = S(\rho_1) + S(\rho_2) - S(\rho_{12}),
\end{equation}
of the whole state \(\rho_{12}\), the heat current deficit and total heat current deficit are the same quantities that can be defined for the EPR, viz., the EPR deficit and the total EPR deficit. 
The heat current deficit and the total heat current deficit will help us understand the dynamics of heat flow in the entire system, and how much of it can be looked upon as due to individual intercommunication of the qubits with the bath. 
%
This will help us 
to get an idea of the 
dynamics
of power dissipation of the system to 
the environment. 
We recall a similar exercise performed for work extraction by global and local heat engines \cite{Oppenheim, Horodecki, Sen, Modi, Bera}.





\section{EPR, heat current and heat current deficit in non-Markovian evolution}
\label{sec4}
\begin{figure}
\includegraphics[height=5cm,width=8cm]{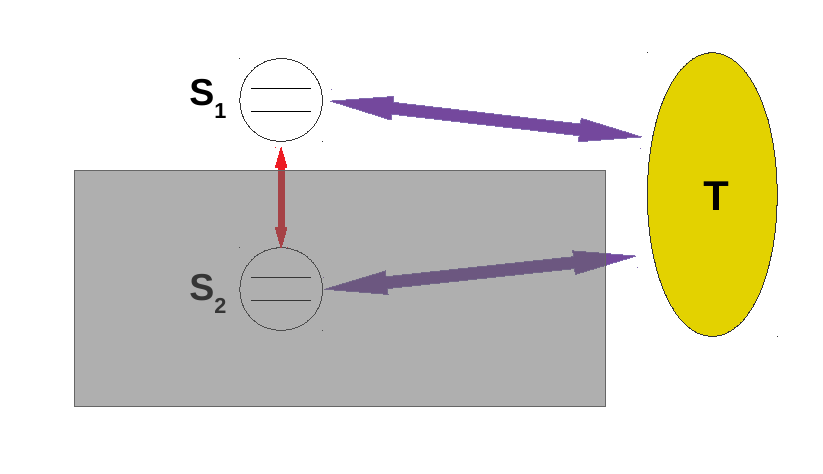}
\caption{Local non-Markovian dynamics. We consider in this case the reduced quantum evolution of one TLS by tracing out the other TLS from the dynamical equation of the entire system of two TLSs. The two TLSs are interacting with a bath at temperature $T$, and this global evolution is assumed to be Markovian.}
\label{fig3}
\end{figure}
The dynamical equation for the two TLSs interacting with a single bath within a Markovian approximation (for the entire system of TLSs) is given in Eq. (\ref{dyn_eq}). 
Just like in the two preceding sections, we are again interested
in investigating the local heat currents and EPRs of the individual TLSs with their respective environments. 
However, instead of working within the lone-qubit Markovian approximation, in which the open quantum evolution of one TLS ignores the presence of the other, we work here by considering the environment of any one TLS to contain the other TLS. 
The scenario is  depicted in Fig. \ref{fig3}. It
typically leads to  a non-Markovian dynamics for the individual TLSs \cite{Jiang, Abiuso, Stegmann} 
and, as we will find, converges to the lone-qubit Markovian limit in the case of no interaction between the two TLSs.  
The dynamical equation for the $i$th system is obtained by taking the trace on the $j$th system of the dynamical equation of the entire system and is given by 
\begin{equation}
\frac{d\tilde{\rho_{i}}}{dt}=\text{Tr}_j \Big(\frac{d\rho_{ij}}{dt}\Big)=\text{Tr}_j [\mathcal{L}_{ij}(\rho_{ij})]\equiv\tilde{\mathcal{L}}_{i}(\tilde{\rho_{i}}),
\end{equation}
where $i \neq j$; \(i,j\in \{1,2\}\); and
$\tilde{\rho}_{i}=\text{Tr}_{j}(\rho_{ij})$. The \(\tilde{\rho}_i\) are also time dependent in general, just like the \(\rho_i\) of the lone-qubit Markovian limit, and the two sets match at \(t=0\) and for any $t$ in the noninteracting case. 

As in Eq. (\ref{hc}), 
the heat current for the $i$th system is defined by
\begin{equation}
\tilde{J}_i=-\frac{1}{T}\text{Tr}[H_{i}\tilde{\mathcal{L}}_{i}(\tilde{\rho_{i}})].
\end{equation}
The relation between the local EPR and local heat current for this non-Markovian situation is given by
\begin{equation}
\frac{dS(\tilde{\rho}_i)}{dt}+\tilde{J}_i = \tilde{\sigma}_i,
\label{EPR_non-marko}
\end{equation}
where $S(\tilde{\rho}_i)=-k_B \text{Tr}[\tilde{\rho}_i\ln(\tilde{\rho}_i)]$. Now, for the non-Markovian evolution considered in this and the succeeding sections, the local EPR for $i$th qubit can be expressed as
\begin{equation}
\tilde{\sigma}_i =-\frac{d}{dt}S(\tilde{\rho}_i(t)||\tilde{\rho}_{{\text{th}}_{i}})-\text{Tr}\Big[\tilde{\mathcal{L}}_{i}(\tilde{\rho}_i)\Big(\frac{H_{i}}{T}+k_B\ln(\tilde{\rho}_{{\text{th}}_{i}})\Big)\Big],
\label{eq:EPR_nm}
\end{equation}
where $\tilde{\rho}_{{\text{th}}_{i}}$ is the state for $i$th system obtained by taking a trace over the other system in the canonical equilibrium state of the entire two-qubit system, i.e., $\text{Tr}_j(\rho_{{\text{th}}_{ij}})=\tilde{\rho}_{{\text{th}}_{i}}$. 
Equation (\ref{eq:EPR_nm}) defines the local EPR for a system evolving in the presence of another, where the duo is undergoing a Markovian evolution. 
If we compare the two equations of local EPRs obtained in this paper, one for the lone-qubit Markovian evolution discussed in preceding sections and given by Eq. (\ref{eq:EPR_m}) and the other being given by Eq. (\ref{eq:EPR_nm}) we first notice the extra term on the right-hand side of Eq. (\ref{eq:EPR_nm}). Furthermore, even the first term of \(\tilde{\sigma}_i\), although structurally similar to the expression for  \(\sigma_i\), is clearly different, as \(\tilde{\rho}_{\text{th}_i}\) is, in general, not \(\rho_{\text{th}_i}\).
We now illustrate these quantities for four paradigmatic Hamiltonians governing the interaction between the two qubits of the system.\par 
%


The dynamical equation can be written as
\begin{equation}
\frac{d\rho_{12}}{d(\tilde{\omega}t)}=\frac{1}{\tilde{\omega}}\mathcal{L}_{12}(\rho_{12})=-\frac{i}{\hbar\tilde{\omega}}[H_{12},\rho_{12}]+\frac{1}{\tilde{\omega}}\mathcal{D}_{12}(\rho_{12})
\end{equation}
so that both  sides 
of the equation are now dimensionless.
For the purpose of depiction in the figures, we will use the dimensionless variable \(\tilde{t} = \tilde{\omega}t\) as the ``time'' with respect to which we will discuss the 
%
natures of the time-dependent quantities $J_i$ and $\sigma_i$, the local heat current and EPR of qubit $i$ in the lone-qubit Markovian limit, and $\tilde{J}_i$ and $\tilde{\sigma}_i$,  the heat current and EPR of qubit $i$ in the non-Markovian case.  All four Hamiltonians considered for the system of two qubits are symmetric with respect to the qubits and we also choose the initial state as symmetric; therefore, it suffices to perform the analysis only for a specific \(i\), and we arbitrarily choose it to be \(i=1\).  
\subsection{Ising interaction between the two TLSs}
\label{A}
\begin{figure*}
\includegraphics[height=6cm,width=18cm]{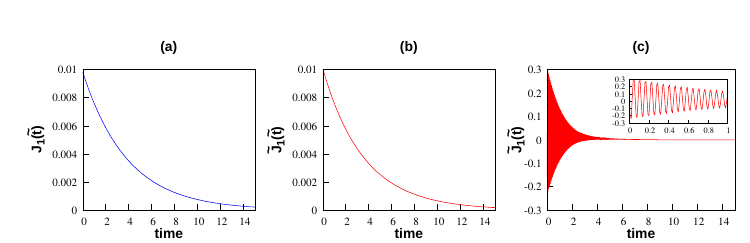}
\caption{Local heat current. We plot here the local heat currents for different cases: (a) the lone-qubit Markovian limit, (b) the Ising interaction, and (c) the \(XYZ\) spin-exchange interaction. The inset in (c) exhibits the same curve as in the main panel but for a shorter time interval to clearly visualize the oscillations of the quantity plotted. 
The initial state of the two qubits in the system is chosen to be
\(\ket{\phi^{+}}=(\ket{00}+\ket{11})/\sqrt{2}\).
For the purpose of the plots,  we have taken \(\omega_1/\tilde{\omega} = 50\), \(\omega_2/\tilde{\omega} = 55\), \(J_z/\tilde{\omega} = 5\) for the Ising interaction case and \(\overline{J}/\tilde{\omega} = 0.8\) for the case of $XYZ$ spin-exchange interaction,
\(k_BT/\hbar\tilde{\omega} = 127.33\) and \(\lambda = 10^{-3}\).
The horizontal axes are in units of \(\tilde{\omega}\), while the vertical ones are in units of \(k_B \tilde{\omega}\). 
}
\label{HC} 
\end{figure*}

\begin{figure*}
\includegraphics[height=6cm,width=18cm]{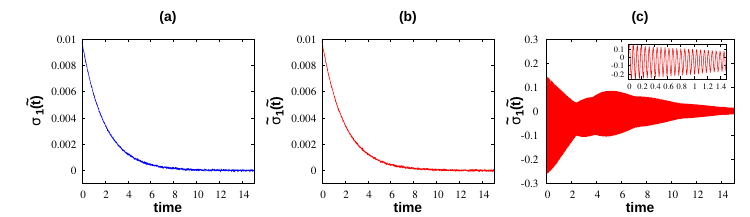}
\caption{Local entropy production rate. We plot here the local EPRs for the same cases for which we plotted the local heat currents in Fig.~\ref{HC}. The parameters are exactly the same as in that figure, except that the vertical axes represent the corresponding local EPRs in units of \(k_B\tilde{\omega}\). 
}
\label{EPR} 
\end{figure*}

Let us begin with the simplest case where two systems are noninteracting, i.e., we choose $g=0$ in Eq. (\ref{H12}). So, there are four possible transition channels with four energy gaps. The Lindblad operators corresponding to the positive energy gaps are
\begin{eqnarray}
\nonumber
\omega=\hbar\omega_1, &&\quad A_{12}(\omega)=|11\rangle\langle 01|+|10\rangle\langle 00|,\\
\omega=\hbar\omega_2, &&\quad A_{12}(\omega)=|11\rangle\langle 10|+|01\rangle\langle 00|.
\end{eqnarray}
Here \(|0\rangle\) and \(|1\rangle\) are the eigenstates of the Pauli \(\sigma_z\) operator. The negative energies are $-\hbar\omega_1$ and $-\hbar\omega_2$ and the corresponding Lindblad operators are obtained from $A_{12}(-\omega)=A_{12}^{\dagger}(\omega)$. 
The evolved state and its physical characteristics are of course obtainable as a limit of the Ising interaction instance considered below. Moreover, the evolved state in the non-interacting case is the same as that in the lone-qubit Markovian limit. 
[For the profiles of $J_1(\tilde{t})$
and  
$\sigma_1(\tilde{t})$, see Figs.~\ref{HC}(a) and~\ref{EPR}(a), respectively.]



We now consider the case of the Ising spin interaction between the two qubits of the system in a parallel field. The Hamiltonian of the system is then  
given by
\begin{equation}
H_{{12}_{\text{Ising}}}=\frac{\hbar \omega_1}{2}\sigma_z^1+\frac{\hbar \omega_2}{2}\sigma_z^2+\hbar J_z\sigma_z^1 \otimes \sigma_z^2,
\end{equation}
\begin{figure*}
\centering
\includegraphics[width=8cm]{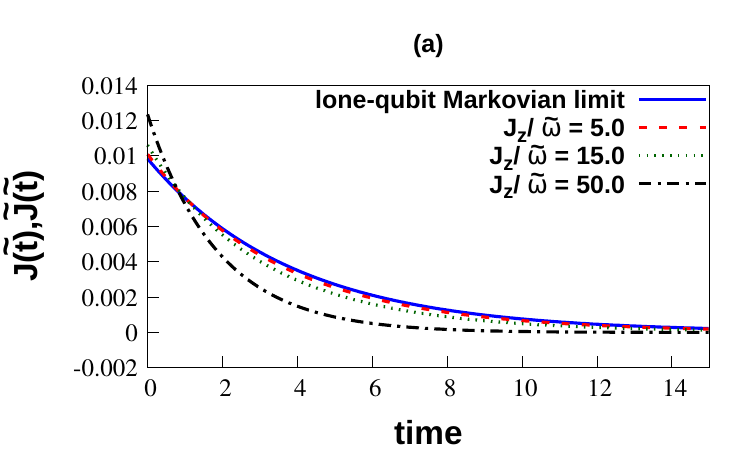}
\hspace{.25cm}%
\includegraphics[width=8cm]{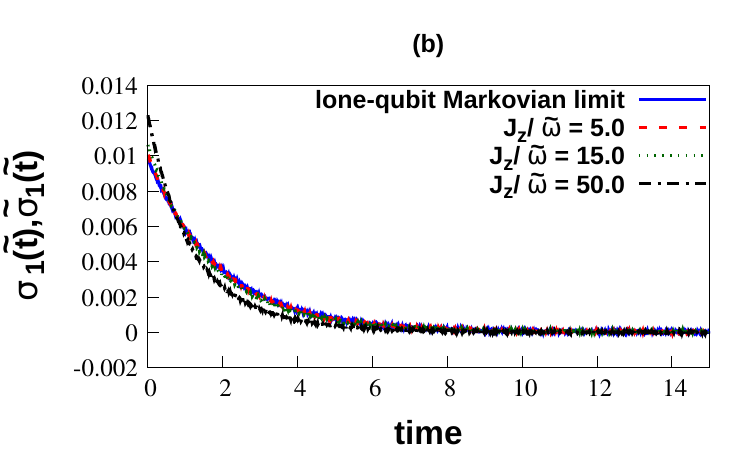}%
\caption{Ising coupling versus lone-qubit Markovian limit. Figures~\ref{HC} and~\ref{EPR} indicate that the heat current and the EPR behave rather similarly for the Ising coupling and the lone-qubit Markovian cases. To investigate this matter further, we compare these quantities in this figure using different coupling strengths of the Ising interaction. The heat currents are plotted in (a), while the EPRs are in (b). All system parameters are the same as in Figs.~\ref{HC} and~\ref{EPR} , except the Ising interaction strengths, which are as noted in the legends. The lone-qubit Markovian limit curves are also plotted in both the panels. The horizontal axes are in units of \(\tilde{\omega}\), while the vertical ones are in units of \(k_B \tilde{\omega}\).}
\label{fig6}
\end{figure*}
where $J_z$ is the Ising coupling strength between the two qubits. Therefore, in Eq. (\ref{H12}) we can choose $g=\hbar J_z$ and \(H_{\text{int}} = \sigma_z^1 \otimes \sigma_z^2\). There are eight possible transition channels and eight possible energy gaps. The positive energy gaps and the corresponding Lindblad operators are
\begin{eqnarray}
\nonumber
\omega=\hbar(\omega_1-2J_z) &&\quad A_{12}(\omega)=|11\rangle\langle 01|,\\\nonumber
\omega=\hbar(\omega_1+2J_z) &&\quad A_{12}(\omega)=|10\rangle\langle 00|,\\\nonumber
\omega=\hbar(\omega_2-2J_z) &&\quad A_{12}(\omega)=|11\rangle\langle 10|,\\
\omega=\hbar(\omega_2+2J_z) &&\quad A_{12}(\omega)=|01\rangle\langle 00|.
\end{eqnarray}
The negative energies are $-\hbar(\omega_1-2J_z), -\hbar(\omega_1+2J_z), -\hbar(\omega_2-2J_z),$ and $-\hbar(\omega_2+2J_z)$ and the corresponding Lindblad operators can be obtained from $A_{12}(-\omega)=A_{12}^{\dagger}(\omega)$. We have depicted the nature of $\tilde{J}_1$ for this case in Fig. \ref{HC}(b). We can see that the natures of $J_1$ [Fig.~\ref{HC}(a)] and $\tilde{J}_1$ [Fig.~\ref{HC}(b)] 
are qualitatively almost the same, although their quantitative values do differ. They are monotonically decreasing with time, finally reaching their steady-state values. The heat currents are positive, implying that heat is flowing from the system to the environment in the transient regime.
The initial state of the system of two qubits for the evolution has been chosen to be $\ket{\phi^+} = (\ket{00} + \ket{11})/\sqrt{2}$.
The behavior of the local heat current for the non-interacting case is qualitatively similar to that in  the Ising case (for nonvanishing interaction strength). The local EPR for the Ising case is presented in Fig. \ref{EPR}(b). 
Just like the local heat currents, \(J_1\) and \(\tilde{J}_1\), the local EPRs, \(\sigma_1\) and \(\tilde{\sigma}_1\), also monotonically decrease, with time, to their steady-state values. While there is qualitative similarity between the heat currents and the EPRs in the lone-qubit Markovian limit and in the Ising interaction case,  the actual quantities differ and depend on the Ising coupling strength 
(see Fig. \ref{fig6}).

\subsection{The $XYZ$ spin-exchange interaction}
\label{C}
If there is an $XY$ spin-exchange interaction between the two qubits of the system along with transverse fields, the Hamiltonian for the system will be
\begin{equation}
H_{{12}_{XY}}=\frac{\hbar \omega_1}{2}\sigma_z^1+\frac{\hbar \omega_2}{2}\sigma_z^2+\hbar (J_x\sigma_x^1 \otimes \sigma_x^2+J_y\sigma_y^1 \otimes \sigma_y^2).
\end{equation}
Here $J_x$ and $J_y$ are coupling constants for interactions in the $x$ and $y$ directions. As is customary, we choose  $J_x = \overline{J}(1 + \delta)$ and $J_y = \overline{J}(1 - \delta )$, with $\delta$ the anisotropy parameter. 
Hence, in Eq. (\ref{H12}) we choose $g=\hbar \overline{J}$ and \(H_{\text{int}} = (1+\delta)\sigma_x^1 \otimes \sigma_x^2+(1-\delta)\sigma_y^1 \otimes \sigma_y^2\).
For the purpose of the plots, we choose 
$\delta = \frac{1}{2}$.
The possible transition energies are $\varepsilon_1-\varepsilon_2, \varepsilon_1+\varepsilon_2,-(\varepsilon_1-\varepsilon_2),$ and $-(\varepsilon_1+\varepsilon_2)$, where
\begin{eqnarray}
\nonumber
\varepsilon_1 &=& \frac{\hbar}{2} \sqrt{(\omega_1 + \omega_2)^2 
+4\overline{J}^2},\\
\varepsilon_2 &=& \frac{\hbar}{2} \sqrt{(\omega_1 - \omega_2)^2
+16\overline{J}^2},
\label{khop-thheke-bero-aj}
\end{eqnarray} 
and the corresponding Lindblad operators can be obtained by using Eq.~(\ref{lindblad}). The natures of the local heat current and local EPR for the $XY$ spin-exchange interaction between the two qubits of the system are qualitatively similar, with the same for the $XYZ$ spin-exchange interaction (discussed below), and so  we have left out the corresponding discussion here.


For an $XYZ$ spin-exchange interaction  present between the two qubits of the system, the Hamiltonian  is
\begin{eqnarray}
\nonumber
H_{{12}_{XYZ}}=\frac{\hbar \omega_1}{2}\sigma_z^1&+&\frac{\hbar \omega_2}{2}\sigma_z^2+\hbar (J_x\sigma_x^1 \otimes \sigma_x^2\\
&+&J_y\sigma_y^1 \otimes \sigma_y^2+J_z\sigma_z^1 \otimes \sigma_z^2),
\end{eqnarray}
where $J_z=\overline{J}$, $J_x = \overline{J}(1 + \delta)$, and $J_y = \overline{J}(1 - \delta )$, so in Eq. (\ref{H12}) \(g=\hbar \overline{J}\) and \(H_{\text{int}} = (1+\delta)\sigma_x^1 \otimes \sigma_x^2
+(1-\delta)\sigma_y^1 \otimes \sigma_y^2+\sigma_z^1 \otimes \sigma_z^2\). 
The possible transition energy gaps are $\varepsilon_1-\varepsilon_4, \varepsilon_2-\varepsilon_3, \varepsilon_1+\varepsilon_3, \varepsilon_2+\varepsilon_4, -(\varepsilon_1-\varepsilon_4), -(\varepsilon_2-\varepsilon_3), -(\varepsilon_1+\varepsilon_3),$ and $-(\varepsilon_2+\varepsilon_4)$, where
\begin{eqnarray}
\nonumber
\varepsilon_{1,2} &=& \mp \overline{J}\hbar+\frac{\hbar}{2}\sqrt{\left(\omega_1 + \omega_2\right)^2
+4\overline{J}^2},\\
\varepsilon_{3,4} &=& \mp \overline{J}\hbar+\frac{\hbar}{2}\sqrt{\left(\omega_1 - \omega_2\right)^2 
+16\overline{J}^2}.
\label{ei-basante}
\end{eqnarray}
The corresponding Lindblad operators are obtained by using Eq. (\ref{lindblad}).
In Fig. \ref{HC}(c) we have depicted the nature of 
$\tilde{J}_1(\tilde{t})$ in the case when there is the  $XYZ$ spin-exchange interaction between the two qubits of the system. We can see that its profile is quite different from the ones in the cases of the Ising interaction and of the lone-qubit Markovian limit.
In the present case, 
$\tilde{J}(\tilde{t})$  changes its sign rapidly in the transient regime and then reaches its steady-state value. Again we have chosen 
the initial state of the two qubits of the system as $\ket{\phi^+}$. In Fig. \ref{EPR}(c) the nature of local EPR in the case of the $XYZ$ spin-exchange interaction 
is presented. The nature of the local EPR is quite similar to that of the local heat current, in that the local EPR also changes sign rapidly in the transient regime before reaching its steady-state value.
\par 
\section{Evidence of Non-Markovianity}
\label{sec5}
The dynamics of the individual qubits studied in the preceding section have been claimed to be non-Markovian, although no definite evidence for the absence of Markovianity was presented. 
We remove that discrepancy in this section by using standard conceptualizations of non-Markovianity. 
\begin{figure*}
\includegraphics[height=14cm,width=18cm]{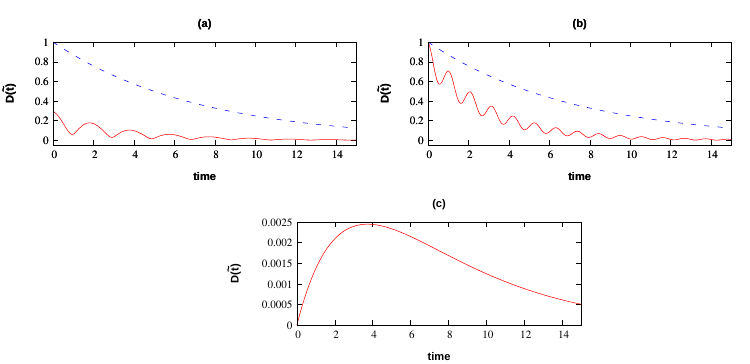}
\caption{The concept of non-Markovianity from Breuer et al.~\cite{Piilo}. We plot here the distance, as a function of time, between evolved states after they start off from certain initial states, for the lone-qubit Markovian limit (blue dashed lines) and the non-Markovian case (red solid lines) for different types of interactions between the two qubits of the system: (a) the Ising interaction case, (b) the case of the $XYZ$ spin-exchange interaction, and (c) non-Markovianity in the case of the $XYZ$ spin-exchange interaction, but for a different initial state. The dimensionless distance is plotted on the vertical axes, while the horizontal axes are the same as in Fig. \ref{HC}: (a) $J_z/\tilde{\omega}=0.8$ and (b) and (c) $\overline{J}/\tilde{\omega}=0.8$. Details about the initial states are given in the main text.
}
\label{BLP} 
\end{figure*}

We begin by using the non-Markovianity measure of Breuer et al.~\cite{Piilo}, which uses the fact that  a Markovian evolution results in reducing the distance (as quantified by certain distance measures, e.g., the trace distance), as time progresses, 
between two (arbitrary) states $\rho(t)=\ket{\psi_t}\bra{\psi_t}$ and $\tau(t)=\ket{\tilde{\psi_t}}\bra{\tilde{\psi_t}}$. The reduction of trace distance between the two states 
signifies the
outflow of information from the system to the environment. Occasional backflow of information is a trait of non-Markovian evolution, which may show up as nonmonotonicity, with time, of the trace distance. The corresponding measure of non-Markovianity has been defined
for the quantum process $\phi(t)$ as
\begin{equation}
N(\phi)=\max_{\rho(0),\tau(0)}\int_{\Lambda > 0} dt \Lambda(t,\rho(0),\tau(0)),
\end{equation}
where $\Lambda(t,\rho(0),\tau(0))=\frac{d}{dt}D(\rho(t),\tau(t))$, with
\begin{equation}
D(\rho(t),\tau(t))=\frac{1}{2}\text{Tr}|\rho(t)-\tau(t)|,
\end{equation}
and where \(\phi(t))(\rho(0) = \rho(t)\) and similarly for \(\tau(t)\). Here $|A|=\sqrt{A^\dagger A}$. The evolution is non-Markovian if $N$ attains a positive value.
In Fig.~\ref{BLP} we provide certain exemplary profiles of the distances (with respect to time) between initial states in the different single-qubit evolutions considered in the preceding section. For the case where the two qubits of the system are interacting with a thermal bath separately and oblivious of the other qubit (lone-qubit Markovian limit), the two initial states are taken as $\ket{\psi_0}=\frac{1}{\sqrt{2}}(\ket{0}+\ket{1})$ and $\ket{\tilde{\psi_0}}=\frac{1}{\sqrt{2}}(\ket{0}-\ket{1})$. The  distance between the corresponding evolved states is depicted as blue dashed lines in Figs. \ref{BLP}(a) and \ref{BLP}(b). 
As expected, the distance falls monotonically with time.  

The other curve in Fig. \ref{BLP}(a) is for the case where the two qubits of the system interact via the Ising interaction (Sec. \ref{A}) and where the state of the two qubits of the system is taken to be 
\(|\psi_0\rangle = (a\ket{00}+b\ket{01}+c\ket{10}+d\ket{11})/\sqrt{N^0}\) and \(|\tilde{\psi_0}\rangle = (\tilde{a}\ket{00}+\tilde{b}\ket{01}+\tilde{c}\ket{10}+\tilde{d}\ket{11})/\sqrt{\tilde{N^0}}\), where the values of the parameters in the states are chosen  independently and Haar uniformly; the particular choices used for the depiction in the red (wiggling) curve in Fig. \ref{BLP}(a) are given by $a=-0.2764+i0.2070$, $\tilde{a}=0.3526+i1.4158$, $b=1.5771+i0.4096$, $\tilde{b}=1.2397-i1.0356$, $c=-0.4097+i0.6886$, $\tilde{c}=0.2556+i1.2417$,
$d=-0.7472-i0.5194$, and $\tilde{d}=-1.1616+i0.2322$. In addition, \(N^0= (|a|^2 + |b|^2 +|c|^2 +|d|^2)^{1/2}\) and \(\tilde{N^0}= (|\tilde{a}|^2 + |\tilde{b}|^2 +|\tilde{c}|^2 +|\tilde{d}|^2)^{1/2}\). The distance depicted with the curve is between the reduced evolved states of the first qubit (\(i=1\)). Note that, unlike in the preceding section, the inputs to the evolution are states that are not symmetric with respect to the qubits of the system. 

The red curve in Fig. \ref{BLP}(b) is for the case of the \(XYZ\) spin-exchange interaction between the qubits of the system (see Sec. \ref{C}). It depicts the distance between the time-evolved states corresponding to the initial states, 
%
$\ket{\psi_0}=\ket{10}$ and $\ket{\tilde{\psi_0}}=\ket{00}$.

The red curves in Figs. \ref{BLP}(a) and \ref{BLP}(b) and  exhibit clear nonmonotonicity of the distance calculated with respect to time, so the measure \(N\) will be positive for these measures. In both these cases, we have considered instances where qubit 1 has different initial states in the evolution. Nonmonotonicity can however appear also if qubit 1 has the same state initially as can be accomplished, e.g., by choosing the two-qubit initial states as $\ket{\psi_0}=\ket{01}$ and $\ket{\tilde{\psi_0}}=\ket{00}$. The interaction between the qubits is chosen as the Ising interaction as in Sec. \ref{A}. In this case, the initial distance between the states of qubit 1 is vanishing. Again, nonmonotonicity shows up (with time) in the distance between the evolved states [see Fig. \ref{BLP}(c)]. 

\begin{figure*}
\includegraphics[height=7cm,width=18cm]{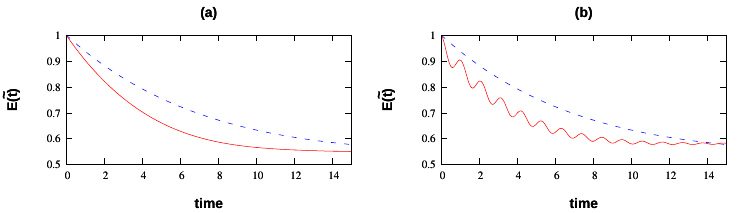}
\caption{The concept of non-Markovianity from Rivas et al~\cite{Huelga}.  We plot here the entanglement, with respect to time, between the system (qubit 1) and an auxiliary, as the two qubits of the system evolve according to different processes dictated by the interactions between the qubits and the environment. The blue dashed curves in both panels are for the lone-qubit Markovian limit, while the red solid curve is for (a) 
the Ising interaction between the qubits for $J_z/\tilde{\omega}=0.8$ and (b) the \(XYZ\) spin-exchange interaction between the qubits for $\overline{J}/\tilde{\omega}=0.8$. 
The horizontal axes are the same as in Fig. \ref{HC}, while the vertical axes are of entanglement (in  units of \(k_B \ln2\)). 
See the text for further details.
%
}
\label{RHP} 
\end{figure*}

A different conceptualization for measuring non-Markovianity was proposed by 
Rivas et al. in Ref. \cite{Huelga}, building on the fact that the system-auxiliary entanglement decays under Markovian dynamics. The corresponding measure 
within a selected time interval $[t_0,t_{\text{max}}]$
is given by
\begin{equation}
I^{(E)}=\int_{t_0}^{t_{\text{max}}} \Big|\frac{dE[\rho_{\text{SA}}(t)]}{dt}\Big|dt-\Delta E.
\end{equation}
Here, $\Delta E= E[\rho_{\text{SA}}(t_0)]-E[\rho_{\text{SA}}(t_{\text{max}})]$ and $E$ is an entanglement measure. The initial system-auxiliary state is taken as the maximally entangled state $|\phi^+_d\rangle=\frac{1}{\sqrt{d}}\sum_{n=0}^{d-1}
\ket{n}
\ket{n}
$. 
For a Markovian evolution, the derivative of entanglement is always negative, as entanglement decreases monotonically in such cases, leading to  $I^{(E)}=0$. A positive \(I^{(E)}\) will imply 
a non-Markovian nature of the evolution. 
In Fig. \ref{RHP} we plot the system-auxiliary entanglement as a function of time for the lone-qubit Markovian limit, the Ising interaction (Sec. \ref{A}), and the \(XYZ\) interaction (Sec. \ref{C}).
For all the curves in Fig. \ref{RHP}, the measure of entanglement is chosen to be the concurrence \cite{Bennett, Hill, Wootters}.
%
%
%
For the lone-qubit  Markovian limit, the system-auxiliary state is 
the 
maximally entangled state, so that  $\rho_{\text{SA}}(0)=\ket{\phi^{+}}\bra{\phi^+}$, and this is shown by the blue dashed lines in Figs. \ref{RHP}(a) and \ref{RHP}(b). 

For the case of the Ising interaction between the system qubits, 
the initial state is $\rho_{\text{SA}}(0)=\rho_{1A}\otimes \rho_{2}$, where 1 and 2 together form the system \(S\), with  $\rho_{1A}$ being the projector on the maximally entangled state $\ket{\phi^{+}}$ and $\rho_{2}=\ket{0}\bra{0}$. The corresponding behavior of entanglement is exhibited as the red curve in Fig. \ref{RHP}(a). The red curve in Fig.~\ref{RHP}(b) exhibits the entanglement dynamics when the interaction is the \(XYZ\) spin-exchange one, with the same initial state as for the Ising interaction. While the \(XYZ\) interaction shows non-monotonicity in the entanglement dynamics, the Ising interaction does not. 


\section{Relations of total heat current deficit in lone-qubit Markovian limit and under non-Markovian dynamics}
\label{sec6}

To better interpret the total heat current deficit, 
we try to find 
a bound on it in this 
section, which also leads us to obtain, in certain cases, a complementarity relation between it and another physical quantity, viz., the entanglement of the two-qubit state. From the balance relation of the entropy production rate and heat current [Eq. (\ref{relation})] and using the definition of the EPR [Eq. (\ref{sigma})], we get
the total heat current deficit 
as
\begin{equation}
\label{delta J}
\int_0^t \Delta J dt=
\chi_{12} - \chi_1 - \chi_2,
\end{equation}
where
\begin{equation}
\label{bhir-kare-imarat}
\chi_a=S(\rho_{a}(0)||\rho_{{\text{th}}_{a}})+S(\rho_{a}(0))
-S(\rho_{a}(t)||\rho_{{\text{th}}_{a}})-S(\rho_{a}(t)),
\end{equation}
with \(a = 12\) or \(1\) or \(2\).



In the lone-qubit Markovian limit, the EPR of the entire system as well as the individual qubits are positive, so that 
$S(\rho(t)||\rho_{\text{th}})\leq S(\rho(0)||\rho_{\text{th}})$. 
As a result,
$S(\rho_{1}(0)||\rho_{{\text{th}}_{1}})-S(\rho_{1}(t)||\rho_{{\text{th}}_{1}})+S(\rho_{2}(0)||\rho_{{\text{th}}_{2}})-S(\rho_{2}(t)||\rho_{{\text{th}}_{2}})\geq 0$, whence
\begin{eqnarray}
\label{bhairab}
\nonumber
\int_0^t \Delta J dt &\leq & [S(\rho_{12}(0)||\rho_{{\text{th}}_{12}})-S(\rho_{12}(t)||\rho_{{\text{th}}_{12}})]\\ &&\phantom{ami to bujhina}- 
I(\rho_{12}(0)) + I(\rho_{12}(t)). \phantom{tum}
\label{UB}
\end{eqnarray}
The first term of the inequality
provides 
the difference of the relative entropy distances of the two-qubit state from the thermal state 
at initial and final (i.e., instantaneous) times. 
The second and third terms are the 
quantum mutual information of the two-qubit state at the initial and instantaneous times, respectively. 
\par
In the case in which the thermal state, \(\rho_{{\text{th}}_{12}}\), is a separable state, that is, it is of the form, 
$\sigma_{12}=\sum_i p_i \sigma_i^1 \otimes \sigma_i^2$, where \(\{p_i\}\) forms a probability distribution and \(\sigma_i^{A,B}\) are density matrices, we have 
\begin{equation}
E_R(\rho_{12}(t))\leq S(\rho_{12}(t)||\rho_{{\text{th}}_{12}}),
\end{equation}
with \(E_R\) denoting the relative entropy of entanglement \cite{Vedral1, Plenio1}, defined for a bipartite state, \(\rho_{12}\), as
\begin{equation}
E_R(\rho_{12})=\min_{\sigma_{12}} S(\rho_{12}||\sigma_{12}),
\end{equation}
where the minimization is over all separable states, $\sigma_{12}$.
In this case, the bound in inequality (\ref{UB}) can be written as a complementarity relation between the total heat current deficit and instantaneous shared relative entropy of entropy of entanglement in the system:
%
\begin{eqnarray}\nonumber
\int_0^t \Delta J dt +E_R(\rho_{12}(t)) \phantom{hasiganmela.................}
\\
\leq   S(\rho_{12}(0)||\rho_{{\text{th}}_{12}}) 
-I(\rho_{12}(0)) + I(\rho_{12}(t)).
\label{LB2}
\end{eqnarray}
A word about the unit of entanglement is in order here. Unlike what is customary in the literature on quantum information science, the logarithms here are chosen to be of base \(e\) and the von Neumann entropy has a multiplicative Boltzmann constant. Therefore, in particular, the entanglement entropy of the (two-qubit) singlet state is \(k_B\ln2\). Let us also mention here that examples of two-qubit Hamiltonians that lead 
to separable thermal states include the ferromagnetic isotropic Heisenberg model and the ferromagnetic and antiferromagnetic $XXZ$ models above certain critical temperatures (see, e.g.,  \cite{Arnesen, Nielsen1, Wang1, Wang2}).

\begin{figure*}
\includegraphics[height=7cm,width=18cm]{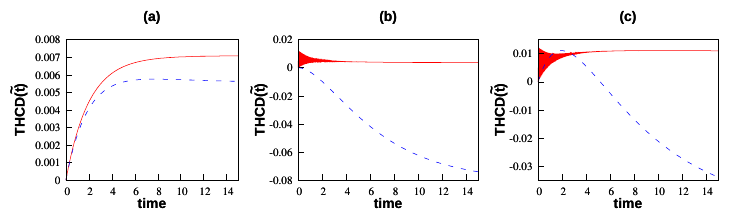}
\caption{Total heat current deficit for the lone-qubit Markovian limit and the non-Markovian case. We plot here the total heat current deficit (THCD) for the lone-qubit Markovian limit, which is the quantity $\int_0^t \Delta J dt$, and represent it by blue dashed lines. The THCD for the non-Markovian cases is $\int_0^t \Delta \tilde{J} dt$ and is depicted here by red solid lines. We show (a) the Ising interaction with $J_z/\tilde{\omega}=0.8$ and (b) the $XY$ and (c) $XYZ$ spin-exchange interactions between the two qubits with $\overline{J}/\tilde{\omega}=0.8$. All other parameters are the same as in Fig.~\ref{HC}. Recall that the quantity \(\chi_{12}\) present in the THCD for the lone-qubit Markovian limit 
depend on the interaction between the two qubits and this leads to the differences between the dashed curves in the three panels. The same \(\chi_{12}\) is also present in the THCD in the non-Markovian case.
The horizontal axes are in units of $\tilde{\omega}$, while the vertical axes are in units of $k_B \tilde{\omega}$.
}
\label{HC_deficit} 
\end{figure*}

We now again look at the total heat current deficit, but in non-Markovian cases. The complete expression for the total heat current deficit can be divided into  Markovian-like and  non-Markovian parts. The Markovian-like part, denoted by \(\int_0^t \Delta \tilde{J} dt |_M\), is given by 
\begin{equation}
\label{mahalagane}
\int_0^t \Delta \tilde{J} dt \Big{|}_M = \chi_{12} - \tilde{\chi}_1 -\tilde{\chi}_2,
\end{equation}
where 
\begin{equation}
\label{gagane}
\tilde{\chi}_i = S(\rho_{i}(0)||\tilde{\rho}_{{\text{th}}_{i}})+S(\rho_{i}(0))
-S(\tilde{\rho}_{i}(t)||\tilde{\rho}_{{\text{th}}_{i}})-S(\tilde{\rho}_{i}(t))
\end{equation}
for \(i=1,2\), where \(\chi_{12}\) is given by Eq.~(\ref{bhir-kare-imarat}).
Note that the Markovian-like part is structurally similar to that of \(\int_0^t\Delta J dt\) in the lone-qubit Markovian limit given in Eq.~(\ref{delta J}). However, there are significant differences, which are indicated by tildes in Eqs.~(\ref{mahalagane}) and (\ref{gagane}). In particular, \(\tilde{\rho}_{\text{th}_i}\) is not necessarily the thermal state of qubit \(i\). Also, the evolved state \(\tilde{\rho}_i(t)\) is obtained by tracing out the other qubit from the two-qubit evolved state \(\rho_{12}(t)\).
Moreover, the second term
\begin{equation}
-\text{Tr}\Big[\tilde{\mathcal{L}}_{i}(\tilde{\rho}_i)\Big(\frac{H_{i}}{T}+k_B\ln(\tilde{\rho}_{{\text{th}}_{i}})\Big)\Big],
\end{equation}
that appears in the non-Markovian case in the expression for \(\tilde{\sigma}_i\) in Eq.~(\ref{eq:EPR_nm}) leads to the non-Markovian part, \(\int_0^t \Delta \tilde{J} dt |_{nM}\), of the total heat current deficit, given by 
\begin{eqnarray}
\label{karnish}
\int_0^t \Delta \tilde{J} dt \Big{|}_{nM} \phantom{ekhan samay holo tomar kachhe apnake}\nonumber \\ 
= \sum_{i=1,2}\text{Tr}\left[\left(\frac{H_i}{T}+k_B\ln\left(\tilde{\rho}_{{\text{th}}_{i}}\right)\right)[\tilde{\rho}_i(t)-\tilde{\rho}_i(0)]\right]. \phantom{sei}
\end{eqnarray}
The final expression for the total heat current is therefore given by 
\begin{equation}
\int_0^t \Delta \tilde{J} dt  = \int_0^t \Delta \tilde{J} dt \Big|_{M}+ \int_0^t \Delta \tilde{J} dt \Big|_{nM},
\end{equation}
with the quantities on the right-hand side given by Eqs.~(\ref{mahalagane}) and (\ref{karnish}), respectively. 

Just like in the lone-qubit Markovian limit, in the case where the two-qubit canonical equilibrium state \(\rho_{\text{th}_{12}}\) is a separable state, we can obtain a complementary relation between the total heat current deficit and the relative entropy of entanglement in the time-evolved state of the two qubits of the system, like in (\ref{LB2}), with extra terms in the bound on the complementarity. Precisely, we have 
\begin{eqnarray}
\label{shaon-raat}
\int_0^t \Delta \tilde{J} dt + E_R(\rho_{12}(t)) \phantom{dighir-ghaTe-anmone-ke} \nonumber \\
\leq S(\rho_{12}(0)|| \rho_{\text{th}_{12}}) -I(\rho_{12}(0)) + I(\rho_{12}(t)) + \mathcal{B}_{nM}.\phantom{eei}
\end{eqnarray}
The part of the complementarity bound [right-hand side of (\ref{shaon-raat})] other than \(\mathcal{B}_{nM}\) is exactly the same as in the  complementarity in the lone-qubit Markovian limit in (\ref{LB2}), and this extra quantity is given by 
\begin{eqnarray}\nonumber
\mathcal{B}_{nM} = -\sum_{i=1,2} \left[S\left(\tilde{\rho}_i(0)||\tilde{\rho}_{\text{th}_i}\right) - S\left(\tilde{\rho}_i(t)||\tilde{\rho}_{\text{th}_i}\right)\right] \phantom{nirjan.upl}\\+ \int_0^t\Delta\tilde{J}dt\Big{|}_{nM}.\phantom{sujan}
\end{eqnarray}
Spohn's theorem rendered the first term of \(\mathcal{B}_{nM}\) as negative in the lone-qubit Markovian limit. The second term of \(\mathcal{B}_{nM}\) did not exist in that limit. 

In the case in which the separability of the two-qubit canonical equilibrium does not hold or is unknown, we have the relation 
\begin{eqnarray}
\label{sujan-majhi}
\nonumber \int_0^t \Delta \tilde{J} dt  \leq
 \left[S(\rho_{12}(0)|| \rho_{\text{th}_{12}}) - S(\rho_{12}(t)|| \rho_{\text{th}_{12}})\right] \phantom{paTatan}\\ -I(\rho_{12}(0)) + I(\rho_{12}(t)) + \mathcal{B}_{nM},\phantom{eei..}
\end{eqnarray}
which is the parallel in the non-Markovian case of the relation (\ref{bhairab}) in the lone-qubit Markovian limit.

Along with being useful in obtaining quantitative estimates of the total heat current deficit, the bounds obtained are important to indicate the relation of a thermodynamic quantity (the deficit) with information-theoretic quantities, such as entanglement and  quantum mutual information. In particular, the bounds have led us to a complementarity between the total heat current deficit and shared entanglement between the two qubits.

The total heat current deficits  for the lone-qubit Markovian limit and in the non-Markovian case for Ising, $XY$ and $XYZ$ spin-exchange interactions are depicted in Fig.~\ref{HC_deficit}. 
For the Ising interaction, we can see that the total heat current deficits for the two cases (lone-qubit Markovian limit and non-Markovian case) are qualitatively similar, whereas for the $XY$ and $XYZ$ spin-exchange interactions, they 
exhibit significantly
different natures. For the $XY$ and $XYZ$ spin-exchange interactions, the total heat current deficits for the lone-qubit Markovian limit are quite similar. For the $XY$ spin-exchange interaction, it decreases from zero, and for the $XYZ$ interaction, it initially increases a little and then decreases. For the non-Markovian case, the total heat current deficits oscillate in the vicinity of zero and then saturate to a positive value for both \(XY\) and \(XYZ\) interactions. For the Ising interaction, the total heat current deficits, for the lone-qubit Markovian limit as well as for the non-Markovian case, initially increase and then reach steady values, with a slight nonmonotonicity in the former case.

\section{Conclusion}
\label{sec7}

We have delved into the important and well-known concepts of heat current and entropy production rate in open quantum evolutions. These quantities appear, for example, in the balance relation for entropy, which is a form of continuity equation for entropy. We inquired about the response of these quantities when we go from a Markovian evolution to a non-Markovian one. We considered a collection of non-Markovian evolutions by looking at the reduced dynamics of a single system between two interacting systems, where the duo is immersed in a heat bath and collectively undergoing a Markovian evolution. The collection of non-Markovian evolutions is generated by choosing several paradigmatic examples of the interaction Hamiltonians acting on the two systems. The Markovian limit of the dynamics is obtained  by choosing to ignore the existence of the other system when examining any system's evolution. It may be noted that the Markovian limit thus obtained is equivalent to the non-Markovian case when the latter's systems are not interacting. We checked the non-Markovianity of the evolutions by using the methods of  Breuer et al. and  Rivas et al. We were interested in understanding the part of the heat current and entropy production rate of the entirety of the two systems (we called them the global quantities) that cannot be accounted for by the same quantities for the individual systems (we called them the local quantities). We performed the analysis for the non-Markovian evolutions as well as for the Markovian evolution for the local dynamics. To this end, we considered and analyzed the differences between the global and the sum of the local quantities and referred them as the heat current deficit and entropy production rate deficit. 
The obtained relations involving these quantities, along with being useful in obtaining their quantitative estimates,   are also important to indicate the connection of a thermodynamic quantity (a deficit) with information-theoretic quantities, such as entanglement and  quantum mutual information.
In particular, we found that the time-integrated heat current deficit, i.e., the total heat current deficit, can in certain instances have a complementary relation with the entanglement between the two systems in the time-evolved state.


\acknowledgements
We acknowledge partial support from the Department of Science and Technology, Government of India, through QuEST Grant No. DST/ICPS/QUST/Theme-3/2019/120.

\end{document}